\documentclass[a4paper,11pt]{article}
\usepackage{jcappub} % for details on the use of the package, please see the JINST-author-manual
\usepackage{lineno}
\title{\boldmath Is dark matter decaying ?}

\author{Jeremy Mould}
\affiliation{Swinburne University,\\
PO Box 218, Hawthorn 3122, Australia}
\emailAdd{jmould@swin.edu.au}

\abstract{An enduring signature of the decay of unstable dark matter constituents into other particles would manifest as a measurable discrepancy in the matter density parameter ($\Omega_{\text{m}}$) between the recombination era ($z \approx 1000$) and the local Universe ($z = 0$). While precision measurements of the Cosmic Microwave Background  tightly constrain the initial matter budget, evaluating this decay hypothesis requires an equally precise audit of the current epoch. We find that current local inventories of baryonic and dark matter are subject to systematic uncertainties - particularly in accounting for the warm-hot intergalactic medium, diffuse intra-cluster media, and the exact profiles of low-mass dark matter halos - rendering a definitive verdict on late-time dark matter decay currently hard to pin down. %out of reach. 
Furthermore, existing astrophysical bounds on cosmic rays, the diffuse gamma-ray background, and reionization history already heavily constrain potential decay channels during this epoch. However, next-generation observational technology is poised to resolve these local accounting gaps. Upcoming high-resolution spectroscopic surveys, next-decade X-ray missions, and advanced weak lensing campaigns will drastically reduce baryon and mass-mapping uncertainties, transforming the late-time matter audit into a cleaner, more definitive test.}

\begin{document}
\maketitle
\flushbottom
\newcommand{\kms}{\mbox{km\,s$^{-1}$}}
\newcommand{\etal}{\mbox{\rm{et al.}~~}}
%\begin{document}
%\centerline{\bf Does dark matter decay ?}
\section{Introduction}
Measurements of $\Omega_{\rm {matter}}$ without qualification are few and far between. $\Omega_m$ (for short) is well determined at redshift z $\sim$
1000 by cosmic microwave background (CMB) measurements (e.g. Planck collaboration 2020; 0.3153 $\pm$ 0.0073),
but measurements at other redshifts rarely reach this high standard. Karachentsev (2026) finds $\Omega_m~\approx$ 0.08
for z $\approx$ 0, and it is possible to integrate the luminosity function and apply a mass to light ratio
at other low %Sloan Digital Sky Survey 
redshifts. Notably,  the  Dark Energy Survey (DES, Popovich \etal 2026; Vincenzi \etal 2024) 
finds $\Omega_m$ = 0.33 and
 $\sigma^{\Lambda CDM}_{\Omega M ,stat+sys}$ = 0.017 in a flat $\Lambda$CDM model from supernova cosmology in the
redshift range (0, 1).
We begin this paper by examining the evidence that the current epoch global density is less than it was before galaxy formation.
We go on to discuss what a reduction with time of the matter density might tell us about the nature of dark matter.
\section{The density at the current epoch}
There are a number of catalogs of nearby galaxies (e.g. Tully 1988; Courtois \etal 2011; CosmicFlows).
The most complete is that of Ohlson \etal (2024) with a redshift distribution in Figure 1.
Galaxies with a negative redshift and no PGC number (Paturel \etal 2003) were edited out.
\begin{figure}
\includegraphics[width=\textwidth]{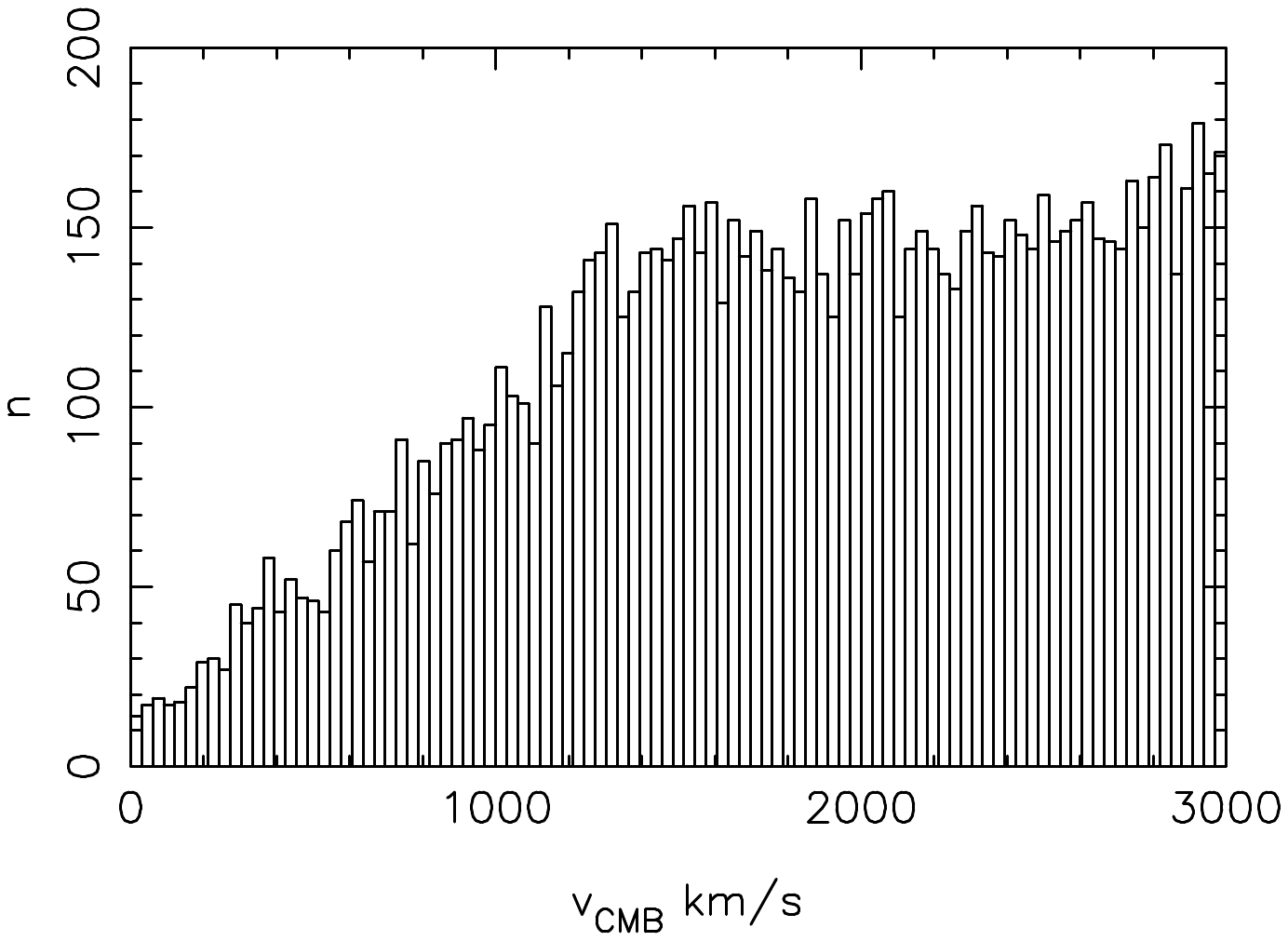}
\caption{Redshifts from the Ohlson \etal (2024) catalog.}
\end{figure}
Figure 1 suggests that incompleteness begins at 1500 \kms, or with the Hubble Constant of Riess \etal (2022) 20.5 Mpc. 
The best available mass measure for nearby galaxies is the 21 cm velocity width w$_{20}$. We adopt a dark matter halo
radius of 100 kpc for all galaxies. In the case of the Milky Way this gives a total mass of 1.0 $\times$ 10$^{12}$ M$_\odot$,
which is consistent with a detailed dynamical model (Binney \& Vasiliev 2020). The average difference between the stellar
mass of Ohlson \etal and the total mass from the CosmicFlows velocity width is 1.78 $\pm$ 0.0175 dex. Although this is a good average, it is mass dependent, and we adopt
the z = 0 dependence given by  Behroozi \etal (2019) and its uncertainty +0.15 -0.03 dex. This catalog comparison also gives us a stellar mass - gas mass difference of 0.53 $\pm$ 0.02 dex.

For galaxies
without a stellar mass we adopted an average total mass from a morphological type - stellar mass relation which falls to 4 $\times$ 10$^7$ M$_\odot$ at type 10. The Virgo and Fornax clusters lie within the fiducial 20.5 Mpc volume,
and their contents exceed the sum of their galaxies. We adopt their masses from Urban \etal
(2011) and    Drinkwater, Gregg \& Colless (2001) and remove catalog galaxies within 4$^\circ$
of their centres. 

We can now begin to populate Table 1 with estimates of $\Omega_m$. We adopt a Hubble Constant of H$_0$ = 73 \kms Mpc$^{-1}$ (Riess \etal 2022), because it is independent of the ionization history of the Universe (Batten \& Mould 2026).
\begin{table}[h]
	\caption{Estimates of density at the current epoch compared with the CMB}
    \vspace{10pt}
	\begin{tabular}{llrrrl}
    \hline
	&	Source      &$\Omega_m$&&&Reference\\
		\hline
	1&	20 Mpc catalog & 0.129 & +0.060 & -0.013&Ohlson+\\
	2&	+ IGM gas      & 0.020 & +0.031&-0.014&Connor+\\
	3&	+ IGM DM       & 0.080  &+0.07&-0.03 &**\\
4&		Total          & 0.228  & +0.072&-0.040\\
%5&		GAMA DR4 M$_*$    & 2.26 $\times$ 10$^{-3}$&$\pm$&0.03 $\times$ 10$^{-3}$&Driver+\\ 
5&		GAMA DR4 M$_*$    & 2.1 $\times$ 10$^{-3}$&$\pm$&0.0006 &Driver+\\ 
6&		10 Mpc catalog M$_*$&2.9 $\times$ 10$^{-3}$       &       &      &Tully+\\
7&		GAMA + gas + DM             & 0.176  &+0.07 & -0.01&Behroozi+\\ 
8&		CMB            & 0.315 &$\pm$&  0.007           &Planck\\
%9&		CMB            & 0.269 &$\pm$&  0.006           &h = 0.73\\
		\hline
		\multicolumn{6}{l}{**Multiplying IGM gas by Planck $\Omega_c/\Omega_b$; ~~DM = dark matter}
	\end{tabular}
\end{table}
Since the discovery of Fast Radio Bursts (FRBs) it has been possible to measure the electron
density in the Intergalactic Medium (IGM) from the dispersion measure of their signal
after correction for the Galactic dispersion measure. Connor \etal (2025) find
that the fraction of the baryons that are in the IGM is 0.76 $\pm$ 0.1. We can go on from there to assume that the IGM baryons are accompanied by their pristine ratio of DM. That is
the third row of Table 1, but it is an overestimate in as much as some of the baryons
are the circumgalactic feedback from star formation and active galactic nuclei. Adding the first 3 rows we get a total $\Omega_m$ from this catalog in row 4.

Some guidance on the IGM fraction of $\Omega_m$ is available from simulations. Shattow \& Croton (2015)
review halo finders applied to n-body runs.
According to SUBFIND, 49\% of the particles are
considered bound at z = 0; %and 49% at z = 0. U
using AHF
(Knollmann \& Knebe 2009), an adaptive mesh halo finder,
38\% of the particles are bound at z = 0. With Rockstar
(Behroozi \etal 2013), a friends-of-friends algorithm with
an adaptive linking length, 55\% of the particles are bound
by z = 0. This would suggest
multiplying line 1 by 1.8 to 2.6 to obtain a total.

An alternative sample to this catalog is the GAMA survey DR4 (Driver \etal 2022). Their
measurement of stellar mass $\Omega_{M*}$ is also shown in Table 1, including their allowance for systematic error and cosmic variance, together with a 30\%
higher density by Tully, Kourchi and Neill (2026). We use the halo-stellar mass 
relation of Behroozi, Wechsler, Hearin \& Conroy (2019) and the GAMA stellar mass function fit 
to convert this to a stellar + gas + dark matter measurement on line 7 of Table 1. This is to %be increased by 30\% for the gas in galaxies and then 
be compared with line 1. Finally, line 8 %\& 9 
gives the Planck value.%, both for h = 0.674 and h = 0.73.
\section{Dark matter decay}
At most Table 1 would support $\delta\Omega_m$ = -0.1 between redshift 1000 and 0, and also possibly zero.
In the first case we consider three distinct possibilities. 
Acharya \&  Johnson (2026) explore a model including decaying dark
matter %(DDMcan alleviate this tension by suppressing the growth of matter fluctuations
%at late times. Specifically, we consider a neutrinophilic 
that has a neutrino decay channel in which a heavy dark
matter particle $\chi$ slowly decays into a Standard Model neutrino and a light invisible fermion,
$\chi \rightarrow  \nu~+~ \phi$. % modifying both the background evolution and the clustering of structure.
% dark matter species that 
The decay is into a lighter daughter particle plus relativistic
 neutrinos. Such decays modify the evolution of the matter density across cosmic time, and the
abundance of clustering DM is reduced at late times. Mould (2026) has drawn attention to something similar, an $increased$ clustering at $early$ times from decaying DM.

A second example of decaying DM is the primordial black hole (PBH). In this case
dm/dt $\propto~-m^{-2}$, and, since this is a power law, we can write it at the half mass point as dlnt/7 = -dlnm/3.
%If m0 is the mass that is just evaporating now at t0, then the mass, m, that was decaying
%at the surface of last photon scattering is given by log m/m0 = --log t/t0 = .
For PBH masses in the asteroid window\footnote{A range allowed by microlensing observations,  $\sim$ 10$^{-17}$
 - 10$^{-11}$ M$_\odot$, according to  Profumo (2026). }, Figure 2 shows that reionization begins far too early 
for m$_{PBH}~<$ 10$^{-13}$ M$_\odot$ ($cf$ Gnedin \& Madau 2022; Nunhokee \etal 2025).
\begin{figure}
	\includegraphics[width=\textwidth]{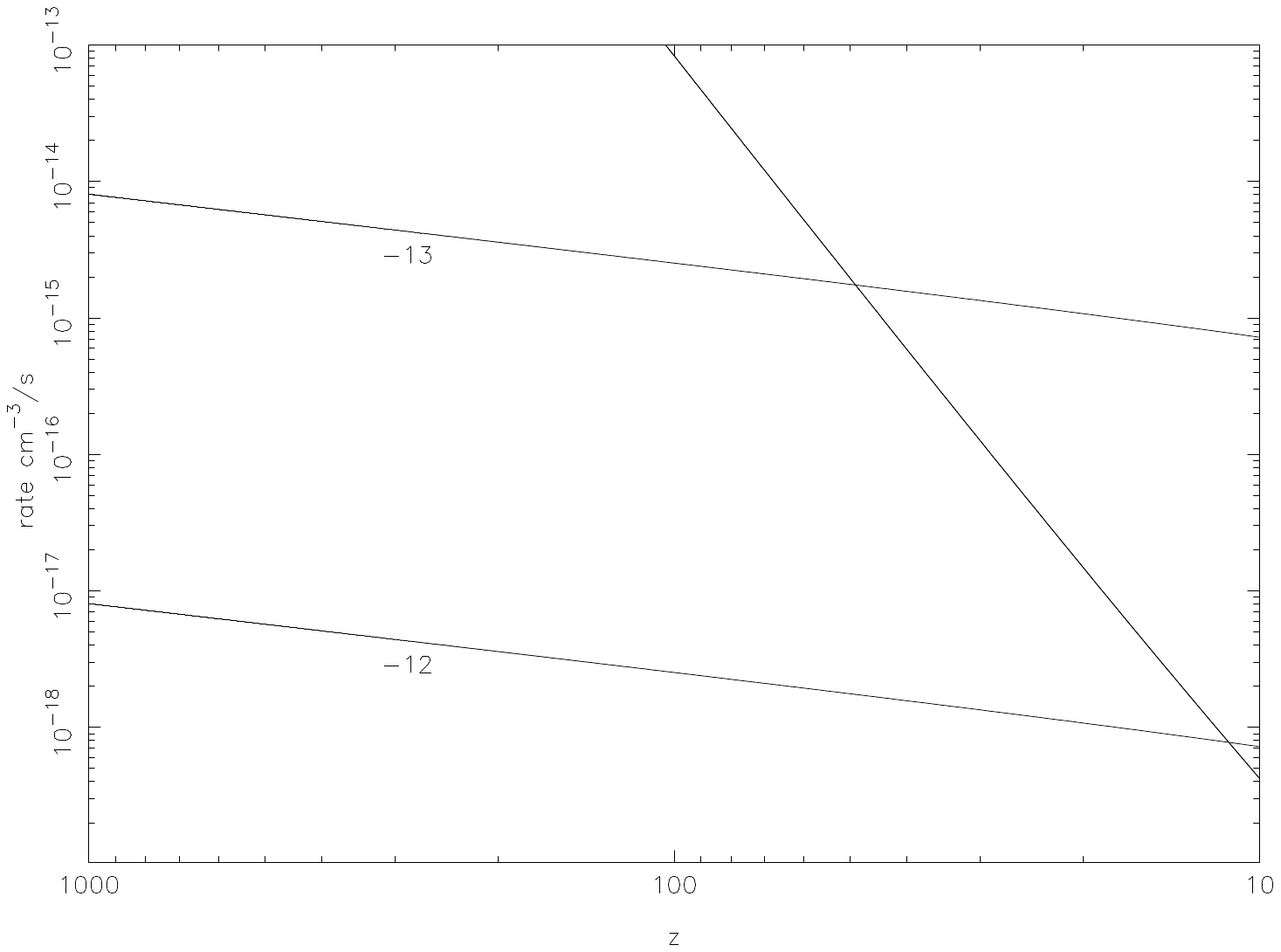}
	\caption{Two PBHs, masses 10$^{-13}$ and  10$^{-12}$ M$_\odot$ have ionization rates
	which intersect the recombination rate of protons and electrons at redshifts 50 and 12
	respectively, assuming they constitute 100\% of the DM. Both are disallowed
	because Lyman $\alpha$ clouds of absorbing neutral hydrogen are observed at z $\sim$ 5.
	The cumulative fraction of their mass lost is very much less than 1\%.}
\end{figure}
%The mass lost depends on the initial mass function (IMF) of PBHs. For an m$^{-1}$ mass function,
%which has equal mass in each decade of mass, the mass lost is      , if the maximum mass
%in the IMF is .
We used the mass loss formalism of Mosbech \& Picker (2022), which for masses over 10$^{15} $ grams is $$\frac{dm}{dt} = -2 \times 10^{-4} \frac{\hbar c^4}{G^2m^2}~~~{\rm gm/s}~~~~~~\eqno(1)$$ and for the recombination rate, with n$_e$ and n$_p$ as the electron and proton density $$\frac{dn(H)}{dt} = 2.7 \times 10^{-13} \surd\frac{10^4 K}{T}n_en_p~~~{\rm cm^{-3}/s}~~~~~~\eqno(2)$$
The density of baryons is given by $$\rho = 1.88 \times 10^{-29} \Omega_b h^2 (1+z)^3~~~   {\rm gm/cc} ~~~\eqno(3)$$ where $\Omega_b h^2$ = 0.0224 according to the Planck collaboration (2020). For DM, $\Omega_C h^2$ = 0.12, and $h$ is the dimensionless Hubble Constant. For 
 m$_{PBH}~>$ 10$^{-12}$ M$_\odot$ an insignificant amount of mass is lost over the age of the Universe.

 One of the hardest components to measure is DM in the IGM. In the case of PBHs the Rubin Observatory will detect millions of supernovae, and microlensing of a supernova at a Gpc
 would result in a 9 hour amplification of the light curve for a 10$^{-9}$ M$_\odot$ lens and 54 minutes for a 10$^{-11}$ one (see Moniez 2000). Only a handful would be expected.

 A third example of DM decay is a finite lifetime of the particle before decay into
 electromagnetic channels. Figure 3 quantifies the rate of decay. $$\frac{dn_X}{dt} = -\frac{n_X}{t_1} \exp(-t/t_1)~~~{\rm cm^{-3}/yr}~~~~~\eqno(4)$$ where n$_X$ is the number density of DM particle X and t$_1$ is its e$^{-1}$ decay time.
\begin{figure}
	\includegraphics[width=\textwidth]{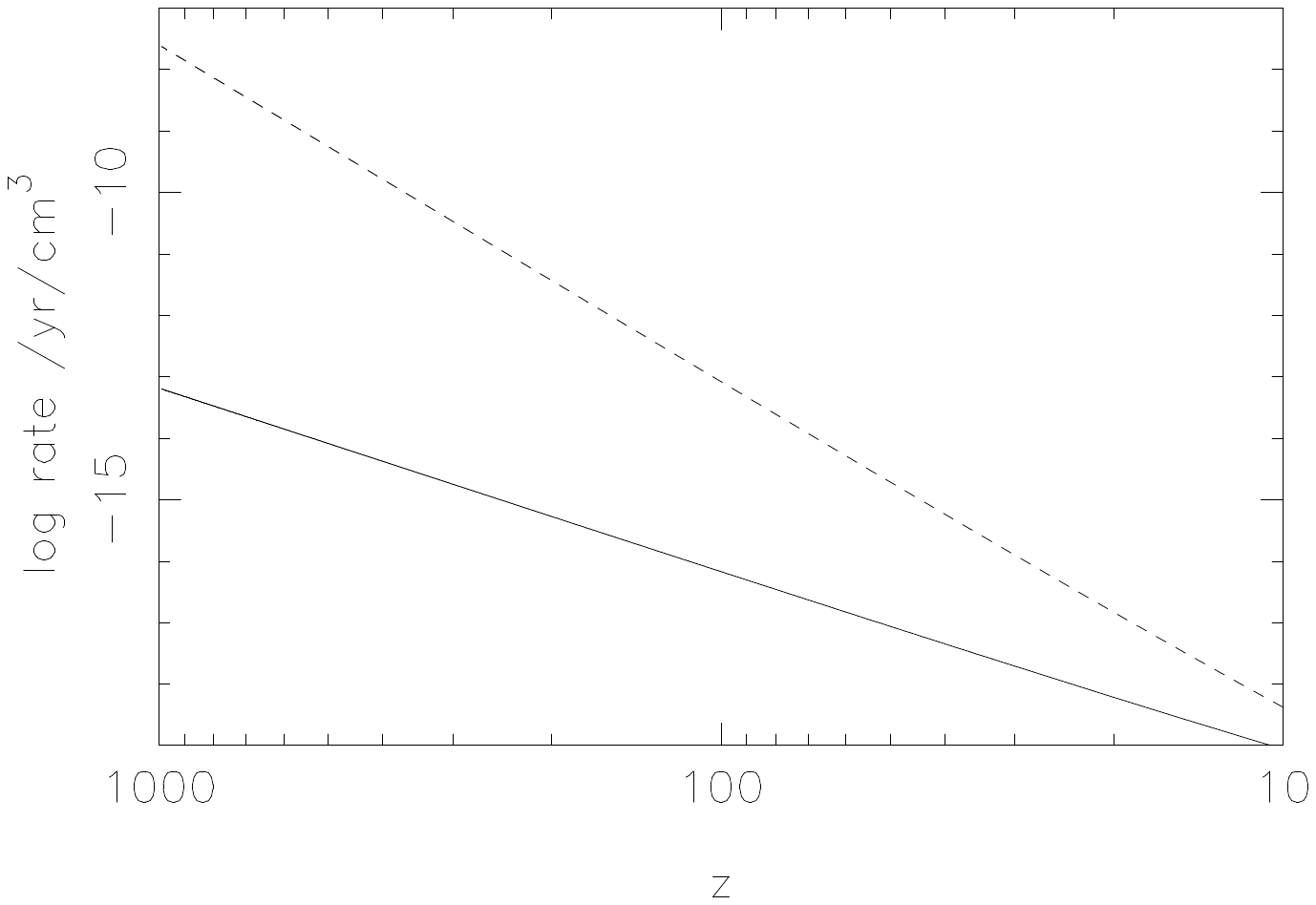}
	\caption{The dashed line is the recombination rate, which must always exceed
	the ionization rate for z $>$ 5 according to observations. An e$^{-1}$ life
	of 1.4 $\times$ 10$^{40}$ years is illustrated here to maintain neutrality over the full range in redshift.
	}
	%	Two PBHs, masses 10$^{-13}$ and  10$^{-12}$ M$_\odot$ have ionization rates
	%which intersect the recombination rate of protons and electrons at redshifts 50 and 12
	%respectively, assuming they constitute 100\% of the DM. Both are disallowed
	%because Lyman $\alpha$ clouds of absorbing neutral hydrogen are observed at z $\sim$ 5.
	%The cumulative fraction of their mass lost is very much less than 1\%.}
\end{figure}
Many channels are open to particle X, and many of these have annihilation radiation as a signature, often considered to contribute to the $Fermi-LAT$ Galactic Centre Excess.
Xu, Qin \& Slatyer (2024) find that while energy injection can significantly increase the temperature of the IGM, from the perspective of CMB anisotropies, the main impact of exotic energy deposition is to raise the global free electron fraction.
Comptonization degrades the decay into photons and other unstable particles, passing all of the
energy into ionization of the baryons. We used Planck collaboration (2020) $\Omega$ values
in this calculation, and a proton mass was the assumed, but barely relevant, DM particle mass.
The DM lifetime is required to exceed estimates of the proton lifetime in this third example.

A further possibility is discussed by  Bencke, Lee \& Kamionkowski (2026), who consider a two-body decay in which a cold DM
 particle $\chi$ with mass m$_\chi$ decays into two identical decay products, each with mass
= $\epsilon m_\chi$/2.
The model is fully characterized by three parameters:
the decay rate, the fraction of the cold DM that is unstable, and the total mass retention $\epsilon$. Energy is conserved by passing it to kinetic energy of the products, thus evading ionization constraints. The model is motivated by recent measurements with the Dark Energy Spectroscopic Instrument (DESI 2026),
which suggest a late-time matter density approximately 5\% lower than that inferred from Planck.
\section{Supernova cosmology}
The supernova standard candle allows us to test the behaviour of $\Omega_m$ over the last 5 Gyrs.
Figure 4 is a fit of the DES % Dark Energy Survey 
supernova data (Popovic \etal 2026; Vincenzi \etal 2024) with 
$\Omega_m$ = 0.1
and an open universe with $\Omega_\Lambda$ = 0.7. The conventional fit is $\Omega_m$ = 0.330 $\pm$ 0.015,
  but the green line is a reasonable fit, given the systematic
uncertainties in both density parameters and the supernova standard candle.
\begin{figure}[h]
\includegraphics[width=\textwidth]{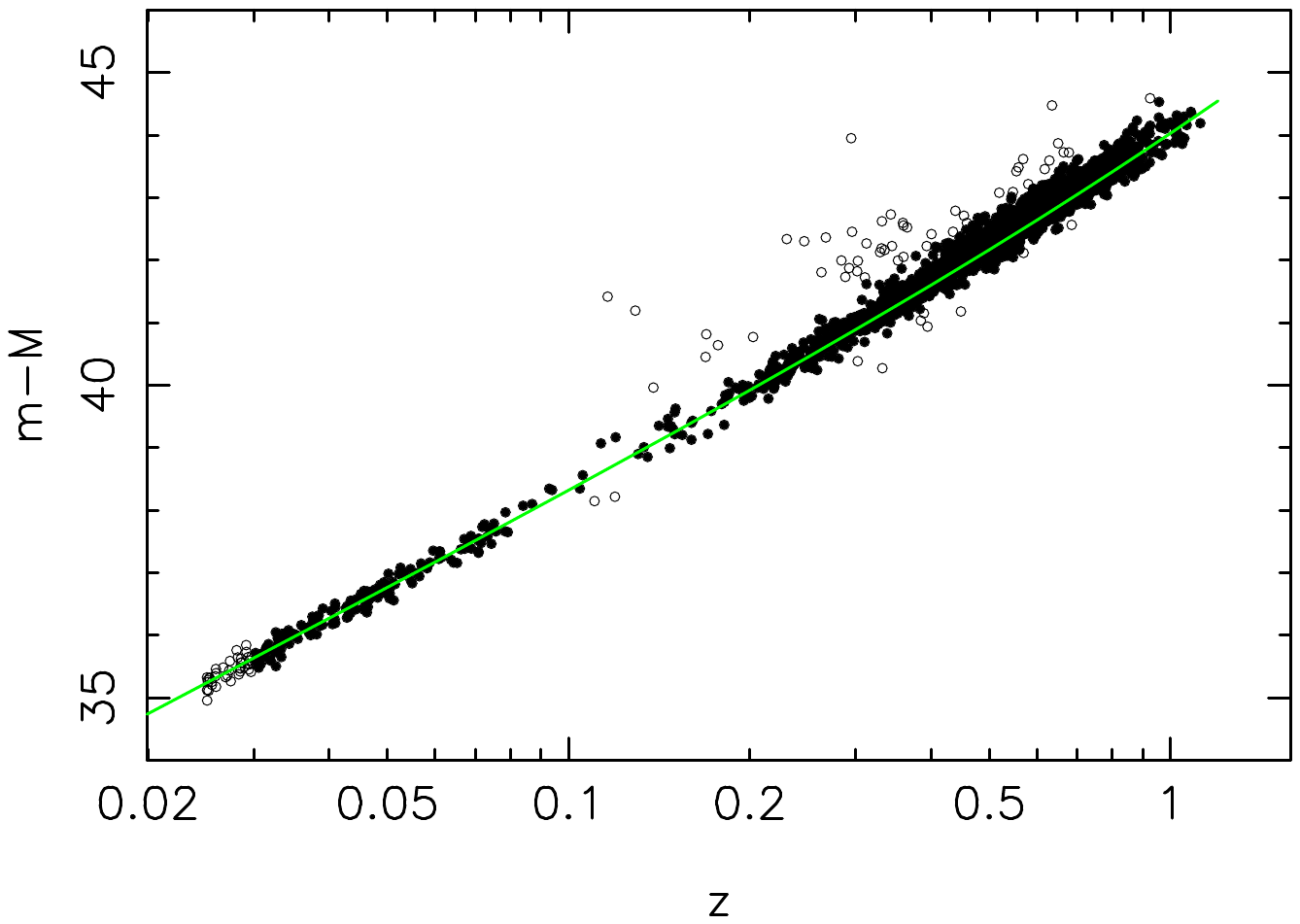}
\caption{The DES supernova distance moduli as a function of redshift. The open circles
are 2$\sigma$ misfits. 
The green line represents $\Omega_m$ = 0.1 and an open universe with $\Omega_\Lambda$ = 0.7.}% It is a reasonable fit, given the
\end{figure}

An alternative view of the same data is by Watkins, Trufax \& Feldman (2026) that the
distance-redshift relation from the local Universe to recombination may already be described remarkably well by a single $\Lambda$CDM model. The combination of CMB, baryon acoustic observations, and supernova observations spans more than four orders of magnitude in redshift and probes the expansion history over nearly the entire age of the Universe. They cite the fact that these measurements can be connected by a single smooth cosmological model is itself a powerful empirical result with the caveat that the remaining discrepancies may arise
from subtle observational systematics or from new physics that preserves the remarkable agreement already seen across such a large range of redshifts. We agree that
future low-redshift supernova measurements, from similar programs designed to match the observational and analysis framework of the forthcoming generation high-redshift surveys, will play a central role in answering that question.

\section{Conclusions}
Galaxy formation from neutral material places significant constraints on DM decay. The
evidence is equivocal that $\Omega_m$ has decreased since the accurate audit available from the CMB. % is equivocal.
Neither catalogs of nearby galaxies nor the supernova standard candle yet clinch the
argument against dark matter decay, although finding a particle/energy dump for lost mass
is very challenging. 

The next decade promises better matter audits at the current epoch,
including Square Kilometer Array 21 cm galaxy mass measurements to a hundred Mpc
and improved knowledge of the IGM density from an all-sky FRB map. 
As noted in the PBH case, it would be highly appropriate for the Vera Rubin Observatory to make a DM detection.
%One of the hardest components to measure is 
Such observations promise to teach us more about
the nature of DM, at least as regards its lifetime.
\section*{References}
Acharya, Y. \& Johnson, R. 2026, arxiv 2606.00529\\
Arnaud, M., Pointecouteau, E. \& Pratt, G. 2005, A\&A, 441, 893\\% 2005, 
Batten, A. \& Mould, J. 2026, submitted to ApJ\\
Behroozi P., Wechsler R. \& Wu H.-Y., 2013, ApJ, 762, 109\\
Behroozi, P., Wechsler, R., Hearin, A. \& Conroy, C. 2019, MNRAS, 488, 3143\\
 Bencke, A. Lee, N. \& Kamionkowski, M. 2026, arxiv 2606.14849\\
Binney, J. \& Vasiliev, E. 2023, MNRAS, 520, 1832\\
Connor, L. \etal 2025, NatAs, 9, 1226\\
%Avinash Chaturvedi1,2, Nicola R. Napolitano3,4 and Michael Hilker 2025\\%Fornax
Courtois, H. \etal 2011, MNRAS, 414, 2005\\
DESI collaboration 2026, arxiv 2404.03002\\
Drinkwater, M., Gregg, M. \& Colless, M. 2001, ApJ, 548, 139\\
Driver, S. \etal 2022, MNRAS, 513, 439\\
%Substructure and Dynamics of the Fornax Cluster
%The 50 Mpc Galaxy Catalog (50 MGC): Consistent and Homogeneous
Gnedin, N. \& Madau, P.
2022, LRCA, 8, 3G\\%2022/12
%Modeling cosmic reionization
Knollmann S., \& Knebe A., 2009, ApJS, 182, 608\\
Moniez, M. 2000, Proc. 35th Rencontres de Moriond, %Cosmological physics with gravitational lensing : 
Les Arcs, France, 2000\\ 
Mosbech, M. \& Picker, Z. 2022, SciPost, 13, 100\\ %Masses, Distances, Colors and Morphologies
Mould, J. 2026, A\&A, 707, 63\\ 
Ohlson, D., Seth, A., Gallo, E.,  Baldassare, V. \& Greene, J. 2024, AJ, 167, 310\\ 
Nunhokee, C. \etal 2025, ApJ, 989, 57\\
Paturel, G. \etal 2003, A\&A, 412, 45\\
Planck collaboration 2020, A\&A, 641, A6\\
Popovic, B. \etal 2026, MNRAS, 548, 632\\%P2026/06
Profumo, S. 2026, arXiv:2606.12775\\
%The Dark Energy Survey supernova program: a reanalysis of cosmology results and evidence for evolving dark energy with an updated Type Ia supernova calibration
%Popovic, B.; Shah, P.; Kenworthy, W. D. and 81 more
Riess, A. \etal 2022, %{\it A comprehensive measurement of the Hubble Constant with 1 \kms/Mpc uncertainty from the Hubble Space Telescope and SHOES team} 
ApJ, 934, 7\\
    %Beyond the halo: redefining environment with unbound matter in N-body simulations
Shattow, G. \& Croton, D. 2015, MNRAS, 452, 1779\\
Tully, R.B. 1988, Nearby Galaxies Catalog, CUP\\
Tully, R., B., Kourkchi, E. \& Neill, J. 2026, arXiv:2606.05780\\
%WISE Photometry of Galaxies within 10 Mpc
Urban, O. \etal 2011, MNRAS, 414, 2101\\
%Urban, O., Werner, N., Simionescu, A., Allen, S. \&  B\"ohringer, H. 
%X-ray spectroscopy of the Virgo Cluster out to the virial radius
Vincenzi, M. \etal 2024, ApJ, 975, 86\\
Watkins, R., Trufax, C. \& Feldman, H. 2026, arxiv 2606.18374\\
Xu, C., Qin, W. \& Slatyer, T. 2024, 
PRD, 110, id.123529\\ 
\acknowledgments
Thanks go to Tamara Davis for our OzDES supernova data and the Australian Research Council for Centre for Dark Matter Particle Physics grant CE200100008.
\end{document}